\documentclass[5p]{elsarticle}

\journal{Nuclear Instruments and Methods in Physics Research, section A}

\begin{document}

\begin{frontmatter}

\title{New readout and data-acquisition system in an electron-tracking Compton camera for MeV gamma-ray astronomy (SMILE-II)}

\author[kyoto_u]{T.~Mizumoto\corref{mycorrespondingauthor}}
\cortext[mycorrespondingauthor]{Corresponding author}
\ead{mizumoto@cr.scphys.kyoto-u.ac.jp}

\author[kyoto_u]{Y.~Matsuoka}
\author[uchu_u,kyoto_u]{Y.~Mizumura}
\author[kyoto_u,uchu_u]{T.~Tanimori}
\author[kyoto_u]{H.~Kubo}
\author[kyoto_u]{A.~Takada}
\author[kyoto_u]{S.~Iwaki}
\author[kyoto_u]{T.~Sawano}
\author[kyoto_u]{K.~Nakamura}
\author[kyoto_u]{S.~Komura}
\author[kyoto_u]{S.~Nakamura}
\author[kyoto_u]{T.~Kishimoto}
\author[kyoto_u]{M.~Oda}
\author[kyoto_u]{S.~Miyamoto}
\author[kyoto_u]{T.~Takemura}
\author[kyoto_u]{J.~D.~Parker}
\author[kyoto_u]{D.~Tomono}
\author[kyoto_u]{S.~Sonoda}
\author[kobe_u]{K.~Miuchi}
\author[tohoku_u]{S.~Kurosawa}

\address[kyoto_u]{Department of Physics, Kyoto University, Kitashirakawa-oiwakecho, Sakyo-ku, Kyoto 606-8502, Japan}
\address[uchu_u]{Unit of Synergetic Studies for Space, Kyoto University, 606-8502 Kyoto, Japan}
\address[kobe_u]{Department of Physics, Kobe University, 658-8501 Kobe, Japan}
\address[tohoku_u]{Institute for Materials Research, Tohoku University, 980-8577 Sendai, Japan}

\begin{abstract}
For MeV gamma-ray astronomy, we have developed an electron-tracking Compton camera (ETCC) as a MeV gamma-ray telescope capable of rejecting the radiation background and attaining the high sensitivity of near 1 mCrab in space. Our ETCC comprises a gaseous time-projection chamber (TPC) with a micro pattern gas detector for tracking recoil electrons and a position-sensitive scintillation camera for detecting scattered gamma rays. After the success of a first balloon experiment in 2006 with a small ETCC (using a 10$\times$10$\times$15 cm$^3$ TPC) for measuring diffuse cosmic and atmospheric sub-MeV gamma rays (Sub-MeV gamma-ray Imaging Loaded-on-balloon Experiment I; SMILE-I), a (30 cm)$^{3}$ medium-sized ETCC was developed to measure MeV gamma-ray spectra from celestial sources, such as the Crab Nebula, with single-day balloon flights (SMILE-II). To achieve this goal, a 100-times-larger detection area compared with that of SMILE-I is required without changing the weight or power consumption of the detector system. In addition, the event rate is also expected to dramatically increase during observation. Here, we describe both the concept and the performance of the new data-acquisition system with this (30 cm)$^{3}$ ETCC to manage 100 times more data while satisfying the severe restrictions regarding the weight and power consumption imposed by a balloon-borne observation. In particular, to improve the detection efficiency of the fine tracks in the TPC from $\sim$10\% to $\sim$100\%, we introduce a new data-handling algorithm in the TPC. Therefore, for efficient management of such large amounts of data, we developed a data-acquisition system with parallel data flow. 
\end{abstract}


\end{frontmatter}

\section{Introduction}
MeV gamma-ray astronomy is unique because nuclear gamma-ray lines in this energy range provide information about fresh isotopes in the universe, such as those produced by nucleosynthesis in supernovae~\cite{nucleosynthesis1,nucleosynthesis2,nucleosynthesis3}, with significant contributions to particle acceleration in active galactic nuclei~\cite{AGN1,AGN2,AGN3,AGN4} or gamma-ray bursts~\cite{GRB1,GRB2,GRB3}, the strong gravity potential of black holes~\cite{BH1,BH2,BH3}, and the interaction of cosmic rays and interstellar matter in the galactic plane~\cite{interstellar_matter1}. However, the observation of such gamma rays in space is difficult because of the large background noise produced by the interaction of cosmic rays with materials around the detector and the atmosphere (albedo). Therefore, MeV gamma-ray astronomy is not as advanced as astronomy in other energy bands. In fact, while EGRET and \textit{Fermi}-LAT found 270 and $\sim$3000 sources over 100 MeV, respectively~\cite{hartman_1999, acero_2015}, COMPTEL loaded on the \textit{Compton Gamma Ray Observatory} (\textit{CGRO}) discovered only about 30 steady gamma-ray sources in all-sky observations~\cite{schonfelder_2000}. Thus, the advancement of MeV gamma-ray astronomy requires better imaging and more efficient background rejection.

\begin{figure}[t]
	\centering
	\includegraphics[width=\linewidth]{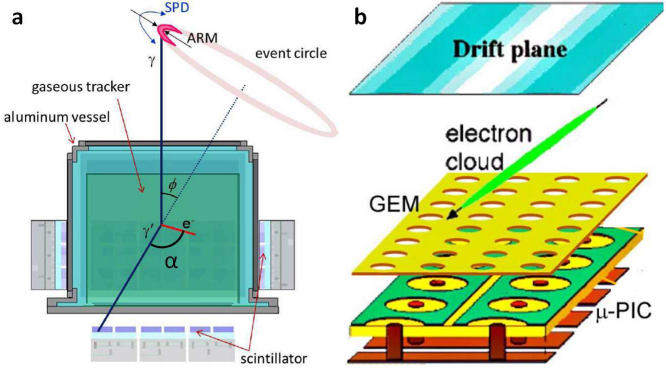}
	\caption{(a) Schematic view of the SMILE-II ETCC system~\cite{sawano_2014}, and (b) the base part of the TPC comprising the $\mu$-PIC and GEM.
		\label{fig:etcc_schematic}
	}
\end{figure}

To address this need, we developed an electron-tracking Compton camera (ETCC), which measures the direction of the recoil electrons in the gaseous time-projection chamber (TPC) event by event. Fig.~\ref{fig:etcc_schematic}a shows a schematic view of the ETCC~\cite{sawano_2014}. By detecting the direction of the recoil electron track, the direction of each incident gamma ray is determined by a point instead of by an event circle. The residual angle $\alpha$ (see Fig.~\ref{fig:etcc_schematic}a) between the scattered gamma ray and the recoil electron provides a robust kinematical background rejection to evaluate the measured angle with that calculated on the basis of the kinematics of Compton scattering. In addition, the energy-loss rate of charged particles in the TPC identifies the recoil particles, which means that the ETCC can select only fully contained electrons, while suppressing escaping electrons, run-through charged particles such as cosmic rays, and neutron-recoil protons. To detect gamma rays from celestial objects, detectors must be deployed at the top of the atmosphere by a balloon or satellite to avoid atmospheric effects. In 2006, we launched the first balloon-borne experiment, called the ``Sub-MeV gamma-ray Imaging Loaded-on-balloon Experiment I (SMILE-I)''. The detector for this mission comprised a small ETCC with a 10$\times$10$\times$15 cm$^{3}$ TPC and 33 pixel scintillator arrays (PSAs) [2112 Gd$_2$SiO$_5$:Ce (GSO) pixel scintillators]. This detector obtained good background noise rejection and observed the spectra of both diffuse cosmic and atmospheric gamma rays in the energy range from 100 keV to 1 MeV~\cite{takada_2011}. In addition, this detector provided good performances for medical applications~\cite{kabuki_2010}.

As a next step, we plan to do a balloon experiment SMILE-II to certify the sufficient capabilities of background rejection and clear imaging for detecting celestial MeV gamma-ray sources by the observation of bright celestial sources such as Crab Nebula or Cygnus X-1. For this purpose, a medium-sized ETCC having approximately 100 times larger effective area than that of SMILE-I with the angular resolution of $<$10 degrees at 662 keV is necessary to observe these targets during single-day balloon flights where the observation time is between 4 and 8 h~\cite{tanimori_2004}. Hence, we developed the medium-sized ETCC with a (30 cm)$^{3}$ TPC and 108 PSAs (6912 GSO pixel scintillators) for SMILE-II\cite{mizumura_2014}. The SMILE-II ETCC also benefited from more efficient tracking of recoil electrons compared with that of SMILE-I: the $\sim$10\% efficiency of SMILE-I~\cite{takada_2011} was improved to $\sim$100\% efficiency by improving the readout electronics and the algorithm for finding tracks in the TPC~\cite{komura_2013}. Together, the larger detection volume and improved tracking method led to the significant increase in effective area, while keeping the weight and power consumption similar to those of SMILE-I. This is important due to the limitations of the balloon experiments. The medium-sized ETCC thus provides more readout channels and generates more data. In addition, the number of triggered events dramatically increases because of the 100-times-greater effective area. As a result, SMILE-II required the redesign of the data acquisition system to handle the 10-times-larger data-transfer rate, all with a similar live-time ratio of $>$70\%. To do this, we improved the entire readout and data acquisition system, including the data-transfer algorithm of the TPC.

This paper first describes the concept of the ETCC and then presents in detail the hardware used for the readout electronics and the algorithm of the new data-acquisition (DAQ) system. Finally, results of observations with SMILE-II ETCC are given.

\section{SMILE-II ETCC}
\begin{figure}[t]
	\centering
	\includegraphics[width=0.9\linewidth]{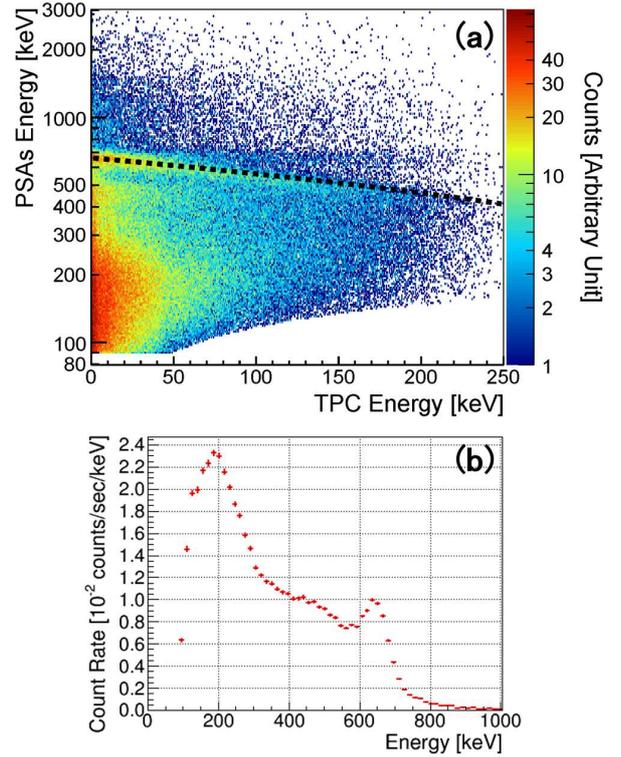}
	\caption{(a) Correlation between energies obtained by PSAs and by TPC upon gamma-ray irradiation from $^{137}$Cs (662 keV). The Compton events are distributed along the dashed line (calculated correlation) with total energy of 662 keV. (b) Energy spectrum measured by combined ETCC.
		\label{fig:etcc_spectra}
	}
\end{figure}
\begin{figure}[t]
	\centering
	\includegraphics[width=0.81\linewidth]{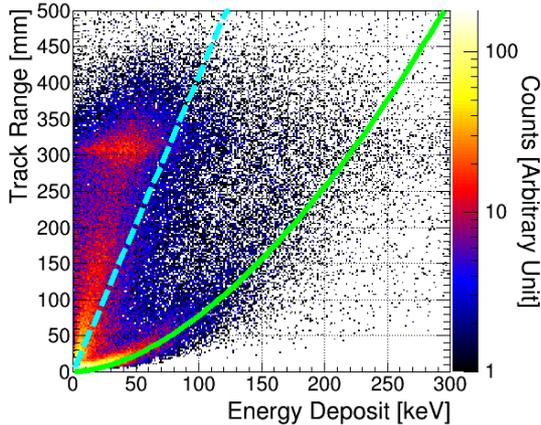}
	\caption{Scatter plot of track range versus energy deposited in the TPC (dE/dx map) due to an intense radiation field created by the irradiation of a water target by 140 MeV protons, where event clusters are clearly grouped into Compton electron stopping in the TPC, minimum-ionizing particles such as cosmic rays and high-energy electrons escaping from the TPC, and protons scattered by fast neutrons. The dashed and solid lines show the energy deposited by minimum-ionizing particles and by fully contained electrons in a 1 atm Ar gas, respectively. The detailed experimental conditions are reported in Ref.~\cite{matsuoka_2015}. 
	  \label{fig:etcc_dEdx}
	}
\end{figure}
\begin{figure}[t]
	\centering
	\includegraphics[width=0.8\linewidth]{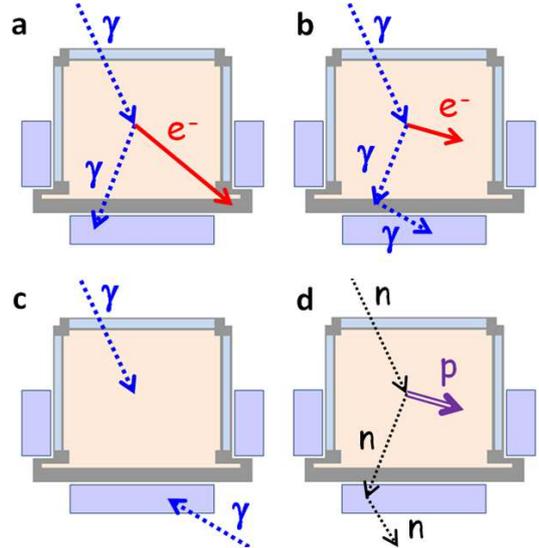}
	\caption{Schematic views of background events in a Compton camera. (a) Electron-escaping event, (b) multiple-scattering event, (c) chance coincident event, and (d) neutron event. An ETCC can reject these background events using Compton kinematics and energy-loss rate in the TPC to determine the angle $\alpha$ and identify the particle, respectively.
	  \label{fig:cc_bg_event}
	}
\end{figure}
An ETCC comprises a three-dimensional gaseous electron tracker for measuring the Compton scattering point, energy, and direction of the recoil electron as well as a position-sensitive scintillation camera constructed from PSAs for measuring the energy and absorption point of the scattered gamma ray. An electron tracker is a TPC based on a micro pixel chamber ($\mu$-PIC)~\cite{ochi_2001}, as shown in Fig.~\ref{fig:etcc_schematic}b. A $\mu$-PIC, which is a two-dimensional micro pattern gaseous detector with pixel electrodes at a density of 400 $\mu$m, has a fine-position resolution of 120 $\mu$m (root-mean square (RMS)) and a high, stable gas gain of $\sim$6000~\cite{nagayoshi_2004}. Since the pixel anodes and cathodes are connected perpendicularly to strip readout electrodes, a 30$\times$30 cm$^2$ $\mu$-PIC has 768 anode strips and 768 cathode strips~\cite{takada_2007}. Since this gain is insufficient for detecting a minimum-ionizing particle, a gas electron multiplier (GEM)~\cite{sauli_1997, tamagawa_2006} is set above the $\mu$-PIC with the low gain operation of about 10, so that this hybrid system provides an adequate and stable gain of approximately 20000 with the gas at normal pressures (Ar 95\%, CF$_4$ 3\%, iso-C$_4$H$_{10}$ 2\%). A hybrid $\mu$-PIC-GEM also reduces ion feedback to the drift electrodes to less than 1\%. To increase the cross section for Compton scattering, Xe or CF$_4$ gas could be used in the ETCC, although this calls for a higher voltage across the electrodes~\cite{takahashi_2011}. In the PSA, we used the GSO scintillator, which is known as a good scintillator material for use in space due to its nonhygroscopicity and radiation hardness. The PSA pixel size is 6$\times$6 mm$^2$ (13 mm in height), and one PSA consists of 8$\times$8 pixels~\cite{nishimura_2007}. This scintillator array was matched to a multianode flat-panel photomultiplier tube (PMT) H8500 (Hamamatsu Photonics), which has 8$\times$8 pixels (6$\times$6 mm$^{2}$ per pixel) and a total geometrical area of 52$\times$52 mm$^2$ (89\% active area). To reduce the number of readout circuits, we used a resistor matrix for each PMT, where the position of the hit pixel was obtained based on the center of the gravity of four signals from the corners of the resistor matrix. The average energy resolution is approximately 11\% at full width at half maximum (FWHM) for 662 keV~\cite{ueno_2012}.

The SMILE-II ETCC shown in Fig.~\ref{fig:etcc_schematic}a uses a (30 cm)$^{3}$ TPC comprising a 30$\times$30 cm$^2$ $\mu$-PIC with a GEM foil stacked above the $\mu$-PIC, and 108 PSAs surround the bottom and four sides of the TPC. Fig.~\ref{fig:etcc_spectra}a shows how the energy obtained by the PSAs correlate with that obtained by the TPC under gamma-ray irradiation from $^{137}$Cs (662 keV) with a calculated correlation (dashed line in this figure), and Fig.~\ref{fig:etcc_spectra}b shows the summed energy spectrum. Fig.~\ref{fig:etcc_dEdx} shows the energy-loss rate (dE/dx) of charged particles in the TPC as a map of the track range versus the energy deposited in the TPC per particle. This plot clearly distinguishes event clusters such as Compton-electron stopping in the TPC, minimum-ionizing particles such as cosmic rays and high-energy electrons escaping from the TPC, and protons scattered by fast neutrons. Thus, the background events depicted in Fig.~\ref{fig:cc_bg_event} can be removed using Compton kinematics to identify the fully contained Compton electron. 

\section{New CMOS ASIC for TPC readout\label{sec:asic}}
\begin{table}[t]
	\centering
	\caption{
		Specifications of FE2009bal.
		\label{tab:fe2009bal}
	}
	\begin{tabular}{cc}
		\hline
		Process	&0.5 $\mu$m CMOS\\
		Number of input	&16 ch\\
		Preamplifier gain	&0.6 V/pC\\
		Peaking time	&30 ns\\
		Sum amplifier gain	&0.8 V/pC\\
		Dynamic range	&$-1$ to $+1$ pC\\
		Cross talk	&$<$ 0.5\%\\
		time walk	&$\sim$6 ns (10 fC to 1 pC)\\
		ENC ($C_d$ = 100 pF)	&$\sim$6000 e$^{-}$\\
		Power consumption	&18 mW/ch\\
		\hline
	\end{tabular}
\end{table}
\begin{figure}[t]
	\centering
	\includegraphics[bb=0 120 950 470,clip,width=\linewidth]{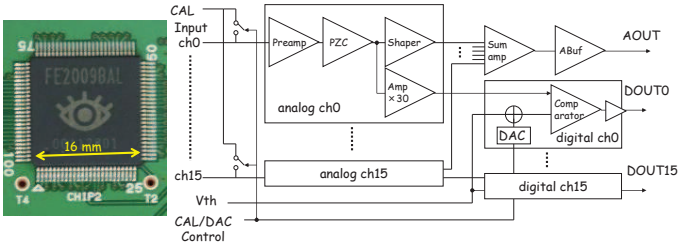}
	\caption{Photograph of fabricated chip (left) and block diagram of FE2009bal CMOS ASIC chips (right).
		\label{fig:fe2009bal_photo}
	}
\end{figure}
\begin{figure}[t]
	\centering
	\includegraphics[width=0.8\linewidth]{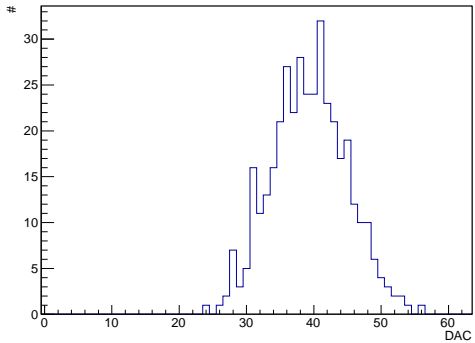}
	\caption{Distribution of the 6-bit DAC for 384 channels in 24 FE2009bal chips.
		\label{fig:fe2009bal_dac}
	}
\end{figure}
\begin{figure}[t]
	\centering
	\includegraphics[width=1.0\linewidth]{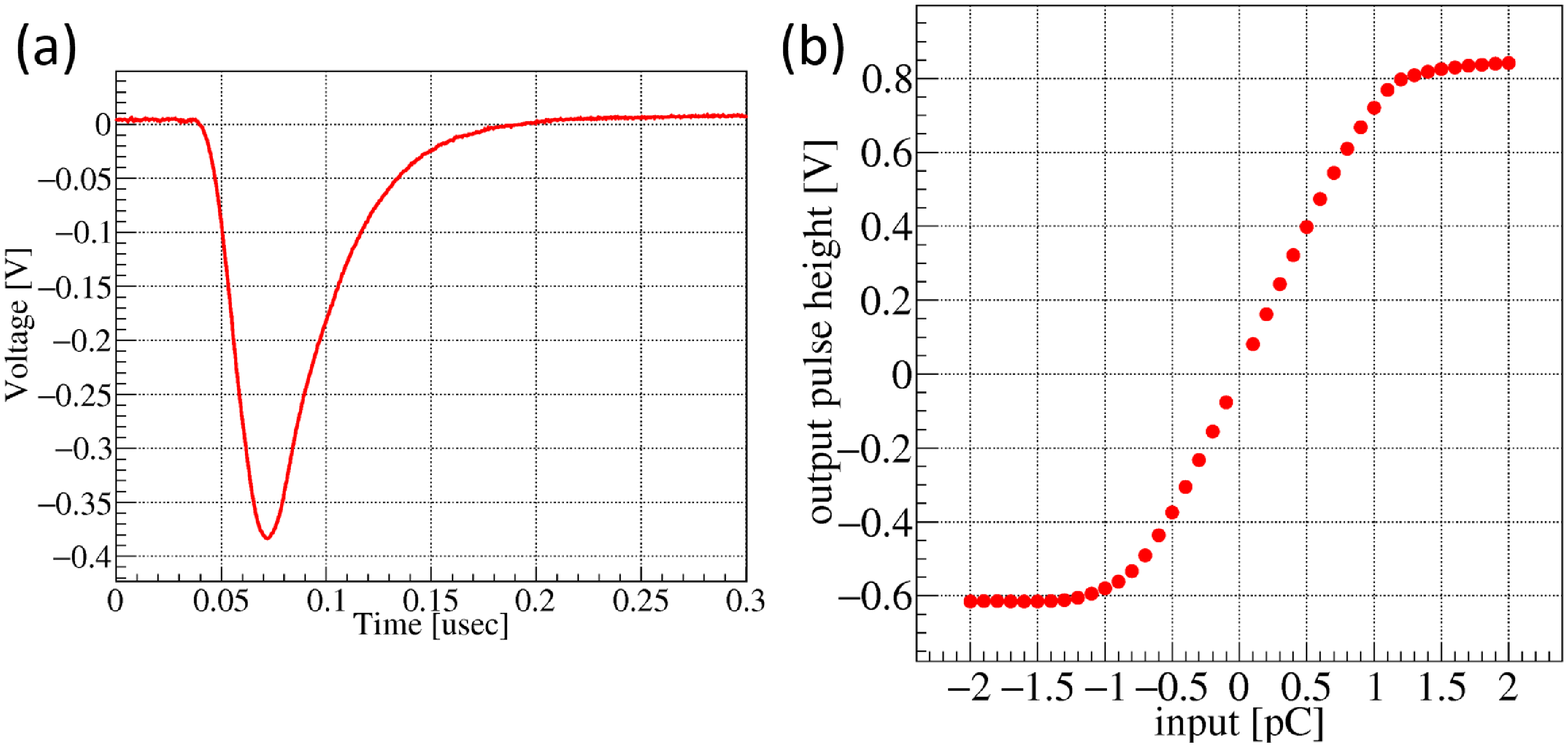}
	\caption{(a) Measured analog sum pulse from FE2009bal. The input charge is -0.5 pC. (b) Measured peak output-pulse voltage of the analog sum pulse from FE2009bal versus input charge.
		\label{fig:fe2009bal_sumamp}
	}
\end{figure}
For the readout electronics for the SMILE-I TPC, we used an amplifier-shaper-discriminator (ASD) integrated circuits that were developed for the ATLAS detector in the large hadron collider~\cite{sasaki_1999} and a position-encoding system constructed from eight field-programmable gate arrays (FPGAs) clocked at 100 MHz~\cite{kubo_2005}. Each strip consumes 130 mW. However, a (30 cm)$^{3}$ TPC with 1536 readout strips is required to keep the total power consumption similar to that of SMILE-I, which corresponds to 40 mW/strip. For this requirement, we developed a new 16$\times$16 mm$^2$ CMOS ASIC (FE2009bal). 

The requirements for the FE2009bal were low power consumption (less than 20 mW/ch, which is a half of the total readout power consumption of 40 mW/ch), high integration (16 ch per chip), wide input dynamic range with both polarities ($-1$ to $+1$ pC), and low noise (equivalent noise charge (ENC) $<$ 6000 e$^-$ at a signal capacitance of 100 pF). To satisfy these requirements, we designed an ASIC chip using the SPICE simulator with a 0.5 $\mu$m CMOS process. The chip was fabricated by the Taiwan Semiconductor Manufacturing Co., Ltd. and contains 16 channels in a 100-pin package (Fig.~\ref{fig:fe2009bal_photo}). The parameters of the manufactured chips are listed in Table~\ref{tab:fe2009bal}. Fig.~\ref{fig:fe2009bal_photo} shows a photograph and a block diagram of a FE2009bal chip, which has 16 input channels, generates a 16-ch-summed analog signal (AOUT), and discriminates CMOS 2.5 V digital outputs for each input (DOUT). The chip requires $\pm$2.5 V power and consumes 18 mW/ch, which satisfies the requirements set out for this chip. The gains of the preamplifier and sum amplifier are measured to be 0.6 and 0.8 V/pC, respectively, and the peaking time of preamplifier is 30 ns. In the comparator, each amplified signal is compared with a common threshold voltage (Vth). Since the input bias voltage of a MOS FET transistor generally varies in each channel, each comparator has a 6-bit digital-to-analog converter (DAC) to compensate for its baseline. Each 6-bit DAC was iteratively adjusted until the counting rate of background noise was equal across all channels, and a serial communication interface was used at each startup to set the data of each DAC. Fig.~\ref{fig:fe2009bal_dac} shows the distribution of the 6-bit DACs for 384 channels in 24 chips. The results show that the DACs have too much variation to use a single DAC value for all strips. Fig.~\ref{fig:fe2009bal_sumamp}a and b shows the analog sum pulse measured from a FE2009bal chip and the measured peak output-pulse voltage versus input charge, respectively. The latter allows us to determine the dynamic range over which the above requirement of linearity is satisfied. In addition, the gain is measured to 0.8 V/pC, which is about twice the gain of the previous ASD chip.

\section{Data acquisition system}
\subsection{TPC readout board}
\begin{figure}[t]
	\centering
	\includegraphics[width=1.0\linewidth]{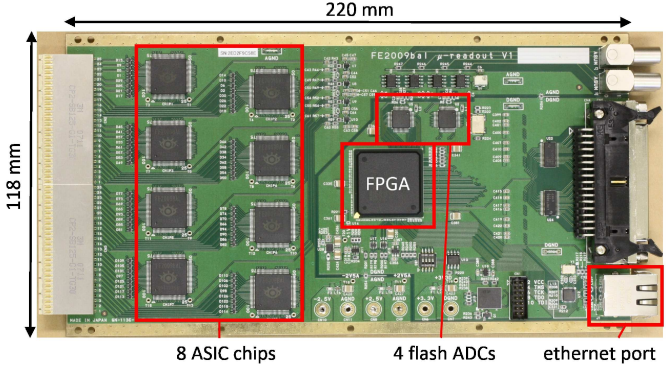}
	\includegraphics[width=1.0\linewidth]{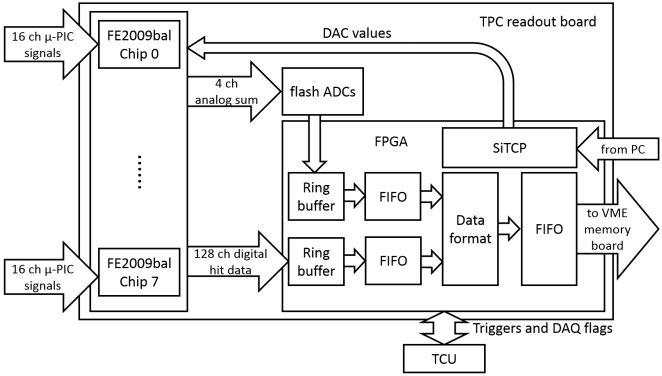}
	\caption{Photograph and block diagram of the new readout boards for the TPC, which contain 8 CMOS ASIC chips (FE2009bal), an FPGA, 4 Flash ADCs, and an Ethernet port.
		\label{fig:tpc_readout_photo}
	}
\end{figure}
\begin{figure}[t]
	\centering
	\includegraphics[width=0.9\linewidth]{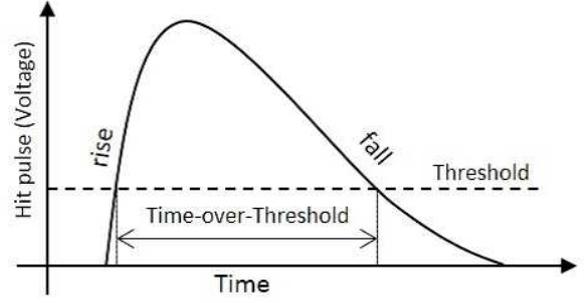}
	\caption{Schematic to define TOT.
		\label{fig:tot_schematic}
	}
\end{figure}
\begin{figure}[t]
	\centering
	\includegraphics[width=0.8\linewidth]{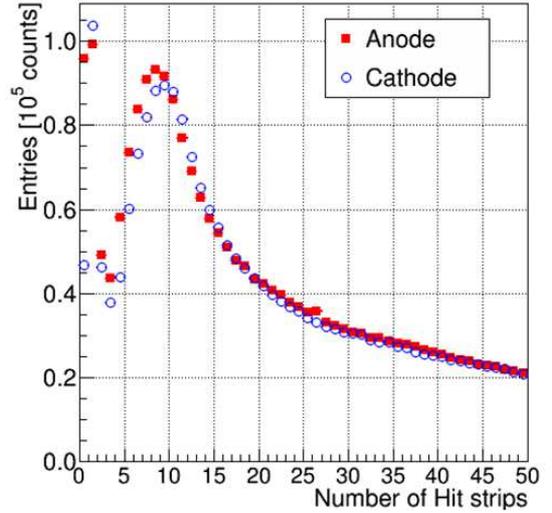}
	\caption{
		Distribution of the number of hit strips of the TPC for recoil electrons for 662 keV gamma rays.
		\label{fig:tot_recoil_e}
	}
\end{figure}
\begin{figure}[t]
	\centering
	\includegraphics[width=\linewidth]{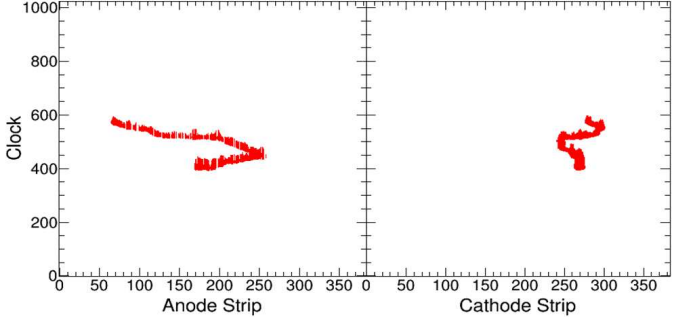}
	\caption{Typical examples of the projected track of a recoil electron, where both hit points are depicted.
		\label{fig:typical_track}
	}
\end{figure}
\begin{figure}[t]
	\centering
	\includegraphics[width=0.8\linewidth]{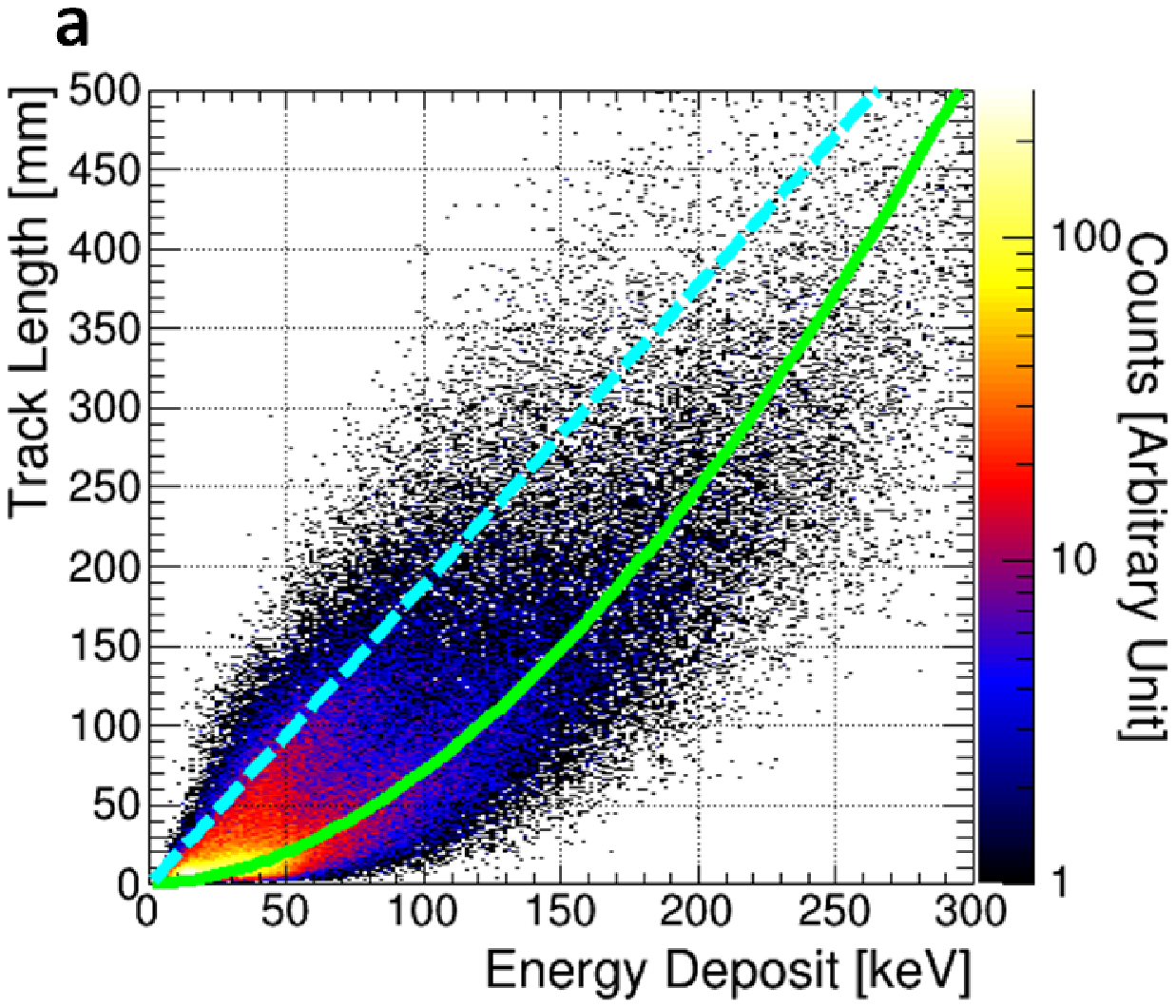}
	\includegraphics[width=0.8\linewidth]{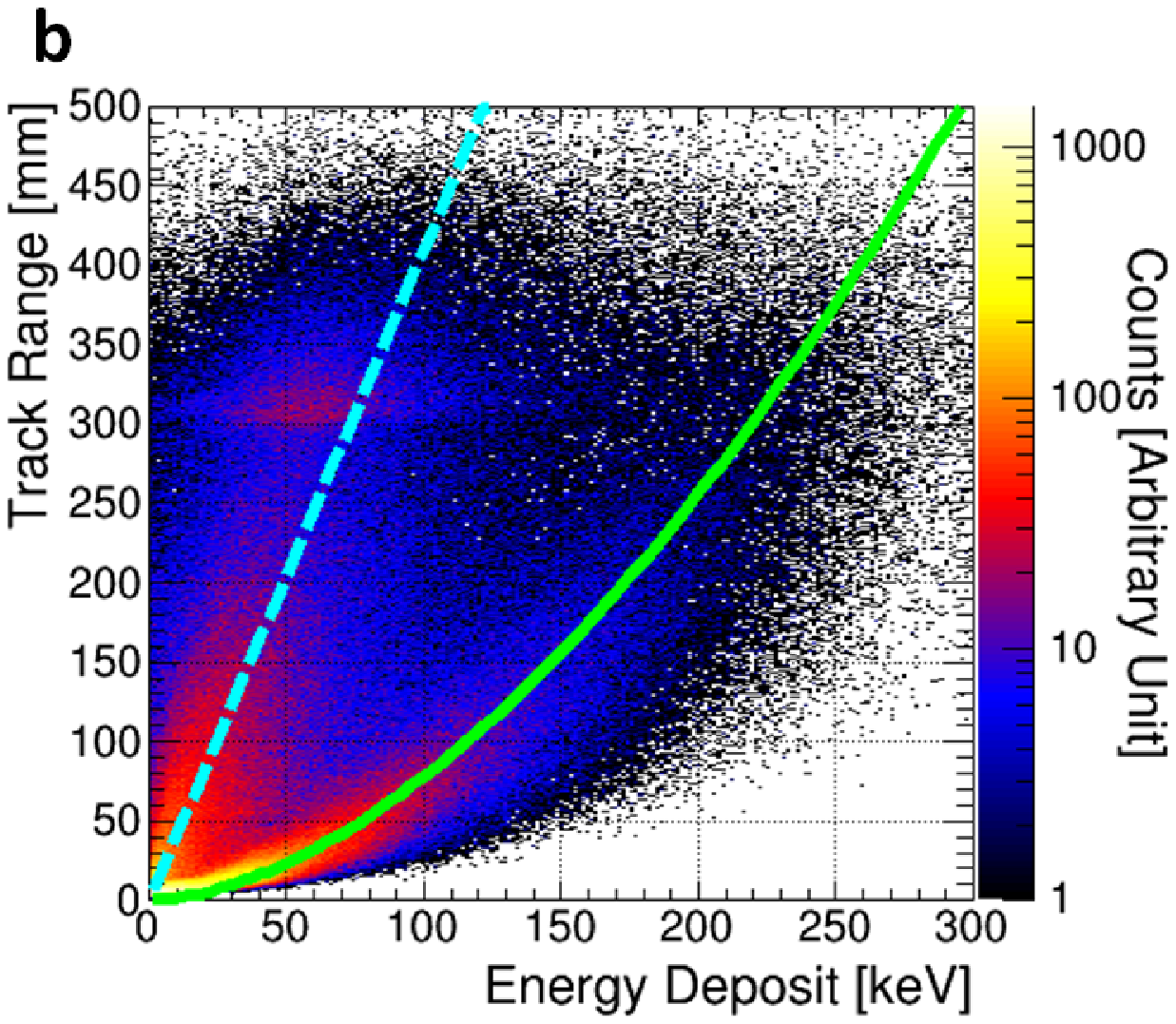}
	\caption{Track length vs the energy deposited for (a) the old (SMILE-I) and (b) new (SMILE-II) reconstruction methods. The dashed and solid lines represent the energy deposited by minimum-ionizing particles and fully-contained electrons, respectively.
	\label{fig:dedx_comp}
	}
\end{figure}
\begin{figure}[t]
	\centering
	\includegraphics[width=0.9\linewidth]{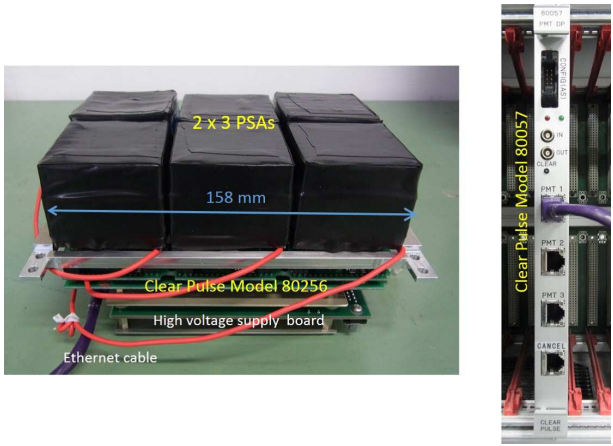}
	\caption{Photographs of readout module. The Clear Pulse Model 80256 is on the left and the VME processor (Clear Pulse Model 80057), which manages four readout modules via Ethernet, is on the right. Six PSAs are mounted on each readout module.
	\label{fig:ha_photo}
	}
\end{figure}
\begin{figure}[t]
	\centering
	\includegraphics[width=\linewidth]{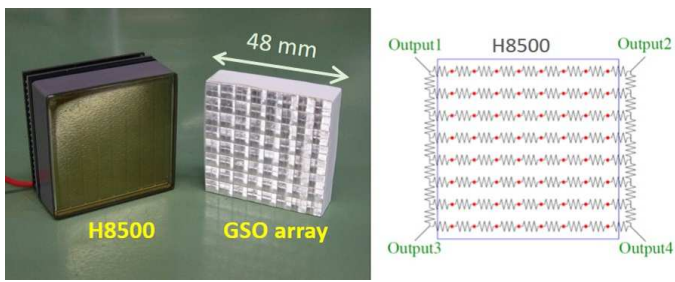}
	\caption{(left) Photograph of PSA components: a H8500 photomultiplier and a GSO crystal array (8$\times$8 pixels). (right) Schematic view of resistor matrix attached to a PSA. Each resistor matrix has 8$\times$8 ch inputs from a PSA (filled circles) and 4 ch outputs.
	\label{fig:psa_photo}
	}
\end{figure}
\begin{figure}[t]
	\centering
	\includegraphics[width=0.8\linewidth]{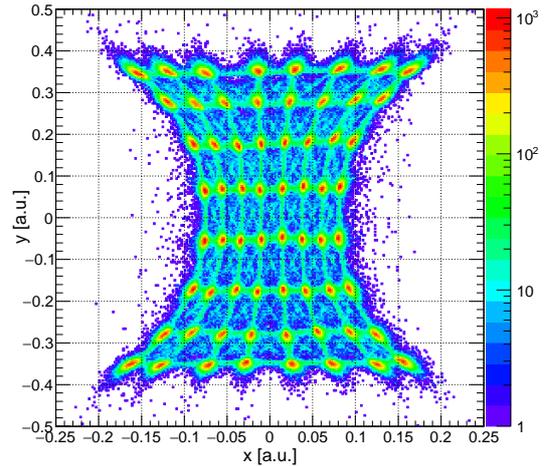}
	\caption{Distribution of hit position in PSA for 662 keV gamma rays. This distribution is derived from the calculation of the center of the gravity of four signals from the corners of the resistor matrix (cf. Fig.~\ref{fig:psa_photo}).
	\label{fig:psa_image}
	}
\end{figure}
\begin{figure}[t]
	\centering
	\includegraphics[width=0.8\linewidth]{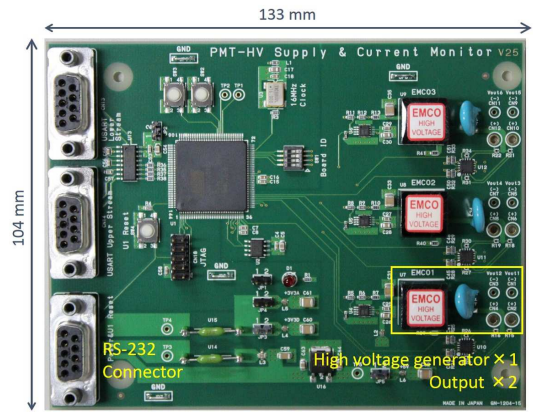}
	\caption{Photograph of high-voltage supply board. Mounted at back of the readout module are three high-voltage generators (each generator supplies the same high voltages to two PSAs), which are controlled via RS-232 by the microprocessor chip on the same board.
	\label{fig:hv_photo}
	}
\end{figure}
\begin{figure}[t]
	\centering
	\includegraphics[width=1.0\linewidth]{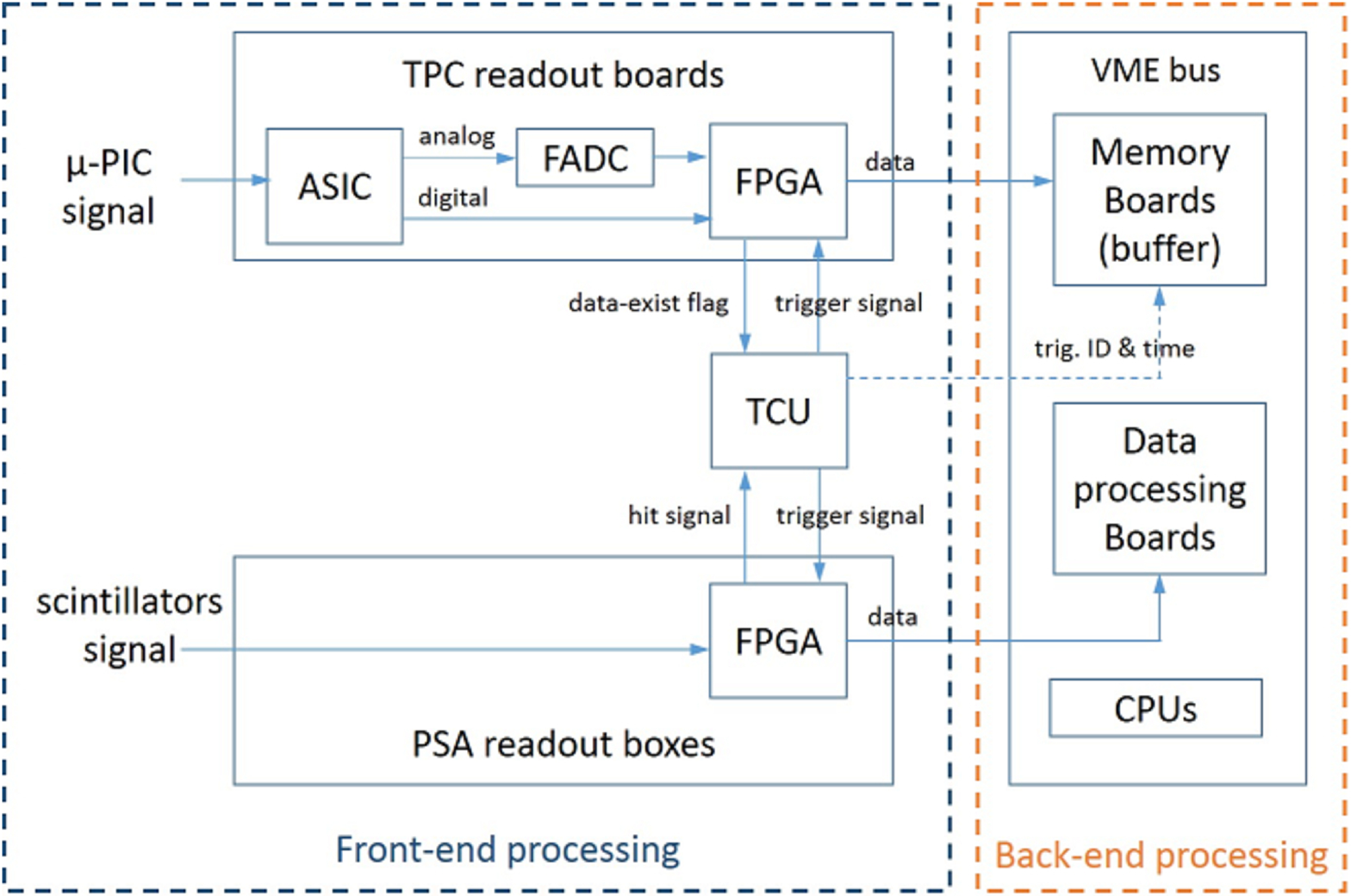}
	\caption{
		Block diagram of SMILE-II DAQ system.
		\label{fig:smile2_block}
	}
\end{figure}
\begin{figure}[t]
	\centering
	\includegraphics[width=0.8\linewidth]{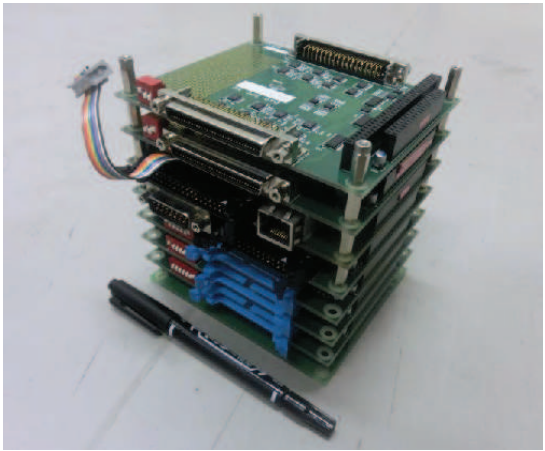}
	\caption{Photograph of TCU, which comprises a main FPGA board, 2 TPC-I/O boards, 3 PSA-I/O boards, a GPS-IO board, and a power board. Each board measures 100 mm $\times$ 120 mm.
		\label{fig:tcu_photo}
	}
\end{figure}
\begin{figure}[t]
	\centering
	\includegraphics[width=0.85\linewidth]{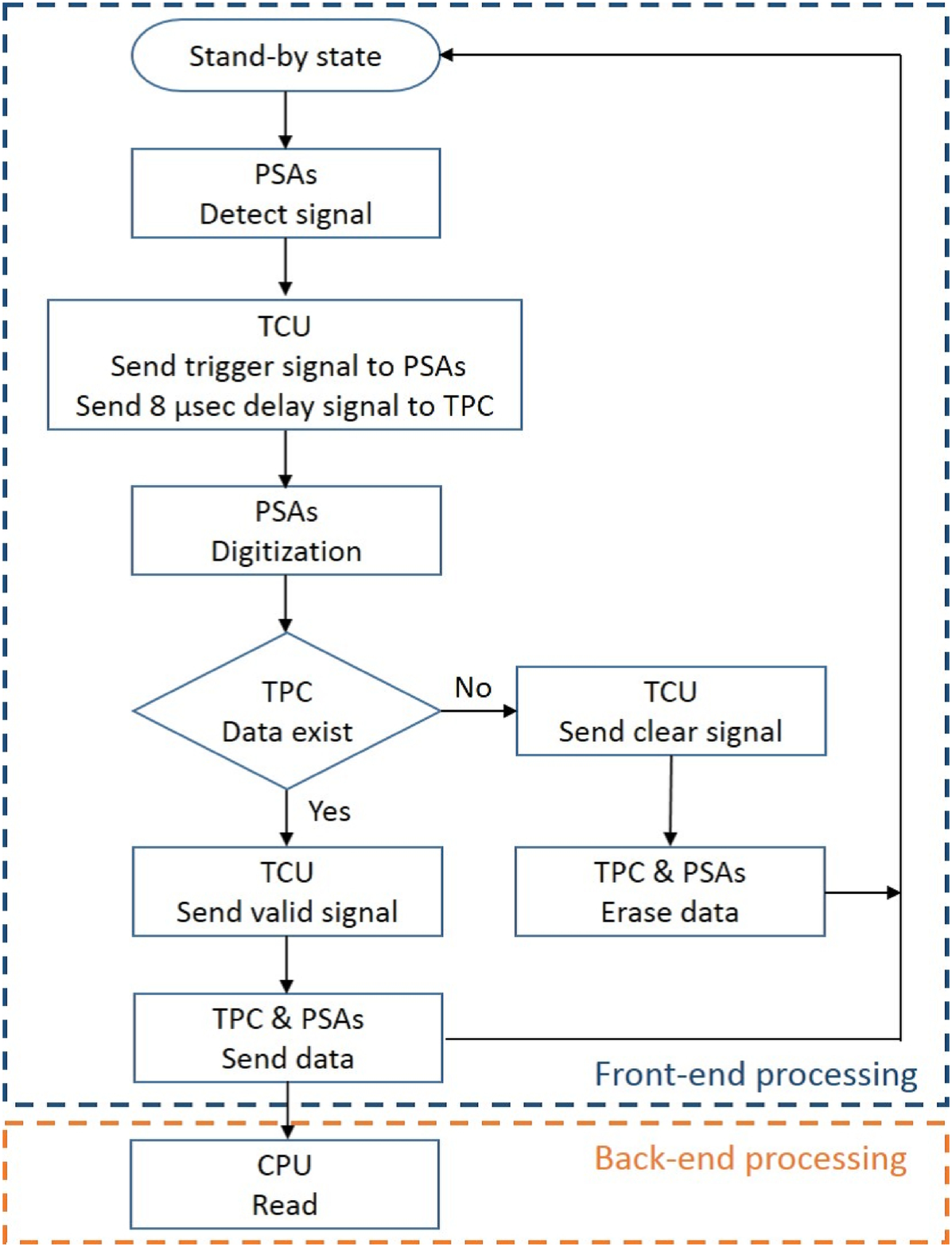}
	\caption{Flow chart of the ETCC mode in the SMILE-II DAQ system. DAQ system mainly comprises 2 parts: front-end system and VME-bus system including main CPU.
	  \label{fig:smile2_flow}
	}
\end{figure}

The new TPC readout board~\cite{mizumoto_2013} contains 4 flash ADCs, an Ethernet port, an FPGA, and 8 CMOS FE2009bal ASIC chips, as shown in Fig.~\ref{fig:tpc_readout_photo}. All discriminated signals from 128 input channels are fed to the FPGA. In the FPGA, the hit patterns of the anode or cathode electrodes are individually synchronized with 100 MHz (10 ns) clocks; hence, the timing resolution of the hit signals of the TPC is determined to be (10 /$\surd$12) $\simeq$3 ns RMS  from the central limit theorem. Four analog signals, each summed over 32 channels, are produced by combining the summed analog output of each pair of adjacent FE2009bal ASIC chips (16 channel-sum $\times 2$). These four signals are then fed to individual 10-bit, 50 MHz flash ADCs, and the resulting digitized data is sent to the FPGA. These ADCs continuously digitize the waveforms of the summed signals at this clock rate. In the FPGA, the 128-bit synchronized hit pattern and 4-ch digitized waveforms are saved to ring buffers, and the 10-bit 50 MHz waveforms are converted into 8-bit 25 MHz waveforms~\cite{matsuoka_2015}. For the determination of the Vth and 6-bit DACs in FE2009bal as mentioned in the section~\ref{sec:asic}, SiTCP~\cite{uchida_2008} for the use of TCP/UDP protocol via the Ethernet is installed in the FPGA. The values obtained for Vth and 6-bit DACs are stored in the main VME CPU and are downloaded via Ethernet to all chips on the readout board at every startup.

Once a trigger signal is input to the FPGA, it stops writing to the ring buffer and starts transferring the formatted data saved over 10 $\mu$s before the trigger signal to a first-in first-out (FIFO) buffer in the FPGA. The process signal is output from the FPGA to the trigger-control unit (TCU), while it reads the ring buffer (the TCU is explained in section 4.3). If the hit pattern exists in the ring buffer, the TPC readout board sends a data-exist flag to the TCU. At the end of the process signal, the TCU orders all TPC readout boards to start data transfer from the FIFO to the VME memory module. It stores data from several thousand triggered events to reduce the dead time due to the latency of the main DAQ-system CPU and due to data transfer from the VME memory. Upon data transfer from the ring buffer to the FIFO buffer, a trigger identification number (trigger ID) is attached to the top of the data. However, if no readout board has hit data, the TCU sends a clear signal to all readout boards. When the TPC readout boards finish data transfer or receive a clear signal, they are all reset and start recording the hits and waveforms of TPC to the ring buffers again. Since the VME memory stores the data from several thousand events, a VME CPU can reduce the number of accesses to the VME memory for data transfer to storage, because the CPU interrupt for accessing the VME system requires a longer latency time (of the order of ms). For SMILE-II, we used an 800 $\mu$m pitch readout that involved grouping two adjacent anode and cathode electrodes; we found that this approach provides a position resolution similar to that of a 400 $\mu$m pitch readout due to the improvement of the signal-to-noise ratio. This strategy results in reduced power consumption and cost. The total power consumption of the readout electronics in the TPC is 45 W (30 mW/strip).

The new TPC readout system was designed to handle the new read-out algorithm for TPC data encoding. The encoding method~\cite{kubo_2005} for the SMILE-I TPC was developed for a $\mu$-PIC high counting X-ray imaging device used in intense synchrotron X-ray sources. In the SMILE-I TPC, when the rising edge of the anode hit coincided within 10 ns to that of a cathode, the addresses and clock counts of the two were automatically encoded and transferred to the CPU. However, a 10 ns gate is too restrictive for the coincidence between the anode and cathode signals of $\mu$-PICs, mainly due to the slow rise time of the amplifiers (16 ns) and to the difference in delay timing in the circuit path of each channel in the FPGA, which caused considerable loss of hit points. In particular, this loss is quite problematic for the ETCC because low-energy recoil electrons of few tens of keV in Compton scattering have only several hit points in the TPC. Actually, the efficiency of $\sim$10\% obtained in the previous ETCC is one order of magnitude less than that calculated from the Compton scattering cross section of the gas in the TPC. To recover almost all the hit points in the TPC, all hit-strip addresses on anodes and cathodes are transferred with the hit timing to the memory module without the coincidence in the encoder. In the off-line analysis, an adequate gate width is applied to anode and cathode hit strips using the hit timing. In addition, as depicted in Fig.~\ref{fig:tot_schematic}, the time between the rise and fall of the hit pulse is recorded as the time-over-threshold (TOT), which is roughly proportional to the pulse height or charge. 

Fig.~\ref{fig:tot_recoil_e} shows the distribution of the number of hit strips of the TPC for recoil electrons due to irradiation of gamma rays from $^{137}$Cs (662 keV). Recoil electrons are clearly separated from the electrical noise. When five hits in one track are required, the electrical noise can be perfectly rejected. Fig.~\ref{fig:typical_track} shows a typical projected track of a recoil electron, where both hit points are depicted. Note that the electron track obtained by the new algorithm indicates the Bragg peak at the track end point by the TOT, reducing misassignment of the scattering point in the previous method due to the loss of hit points. This new reconstruction method provides a much better detection efficiency of $\sim$100\%. Fig.~\ref{fig:dedx_comp}a and b shows how the track range relates to the energy deposited for the old and new reconstruction methods. The new method better identifies recoil electrons that stop in the TPC and separates them from cosmic rays and penetrating high-energy recoil electrons escaping from the TPC, both of which are minimum-ionizing particles. This result is quite consistent with the calculated curves in the figure. Although this new algorithm provides a satisfactory improvement in TPC performance, it also increases the quantity of TPC data, requiring a new TPC readout system with about five-times-faster data transfer capacity than SMILE-I. For this, we adopted the VME memory module for multi-event buffering.

\subsection{PSA readout module}
The PSA-data-handling system comprises a readout module (Clear Pulse Model 80256~\cite{ueno_2012}) and a VME processor (Clear Pulse Model 80057) that manages four readout modules via Ethernet. Six PSAs are mounted on each readout module, which includes 24 preamplifiers, 24 shapers, 24 12-bit ADCs, and 6 comparators, as shown in Fig.~\ref{fig:ha_photo}. A resistor matrix is attached to each PSA, as shown in Fig.~\ref{fig:psa_photo}, and the 8$\times$8 channels from the PSA reduce four signals from the four corners of the resistor matrix. The hit position within a PSA is calculated from the weighted average of four signals. Fig.~\ref{fig:psa_image} shows the hit distribution obtained by this method for 662 keV gamma rays. A threshold for each PSA is set in the readout module via the VME processor module. High voltages for the six PSAs per readout module are supplied by the high-voltage supply board set at the back of the readout module. The photograph of this board is shown in Fig.~\ref{fig:hv_photo}; the three high-voltage generators (one generator supplies the same high voltage to two PSAs) are mounted and controlled via RS-232 by the micro processor chip on the same board.

When a dynode signal from a PSA exceeds the threshold, the readout module sends a hit signal to the TCU and the TCU immediately replies with a start signal to all readout modules to activate the digitization of PSA signals regardless of any coincidence with a hit in the TPC. The threshold is set via the VME processor by the main DAQ CPU. When the TCU receives a data-exists signal from the TPC readout boards, the TCU generate a valid signal for all PSA readout modules for data transfer. If the TPC readout boards have no data, the TCU sends a clear signal to all PSA readout modules, and the PSA readout module is reset and begins to wait for the next PSA hit signal. 

\subsection{Trigger-control unit}
Fig.~\ref{fig:smile2_block} shows a block diagram of the SMILE-II DAQ system, which comprises 2 main parts: the front-end system (TPC, PSA, and TCU) and the VME-bus system (which includes main CPU). Each system works individually. In the front-end system, the TCU shown in Fig.~\ref{fig:tcu_photo} controls starting/stopping of data acquisition and trigger making. The TCU comprises a main board, 2 TPC-input/output (I/O) boards, 3 PSA-I/O boards, GPS-I/O boards, and a power board. The main board has an FPGA and an Ethernet port and is controlled by the DAQ CPU via Ethernet. The FPGA on the main board controls all DAQ signals from the front-end system with a 100 MHz clock. In addition, the DAQ dead time is measured by a clock counter in this FPGA. The TCU main board operates TPC readout boards and PSA readout modules via TPC-I/O boards and PSA-I/O boards, respectively. A TPC-I/O board can communicate with 4 TPC readout boards, and a PSA-I/O board can operate 6 PSA readout modules. The current front-end system is operated with 2 TPC-I/O boards and 3 PSA-I/O boards. Moreover, the TCU offers the option of increasing the number of TPC-I/O and PSA-I/O boards to improve the effective area by incrementing the number of TPC readout boards or PSA readout modules. A GPS-I/O board has a serial input to obtain the global time, a PPS signal input as a force trigger, and 5 veto trigger inputs for charged particle rejection. All boards are powered by the power board.

The TCU has 3 DAQ modes: the ETCC mode, the PSA calibration mode, and the TPC calibration mode. Fig.~\ref{fig:smile2_flow} shows the flow chart of the SMILE-II ETCC mode. In the ETCC mode, if any PSAs in the standby state detect a signal above the threshold, the TCU sends a trigger signal (GSO trigger) to all PSA readout modules via PSA-I/O boards. The trigger for all TPC readout boards is delayed for 8 $\mu$s with respect to the GSO trigger. As mentioned above, the PSA readout modules immediately start the digitization. However, the TPC readout board checks its ring buffer and confirms whether any TPC hit-position data were created within 10 $\mu$s, which is longer than the drift time of 6 $\mu$s for the use of the Ar based gas in the TPC. When the TCU accepts the data-exists flag (cf. section 4.1) from the TPC readout board, the TCU sends a valid signal for all systems in the ETCC. However, if no signal is transmitted from the TPC, the TCU sends the clear signal, and all readout boards erase their data and return to the standby state. Once a valid signal from the TCU is accepted, all TPC readout boards and PSA readout modules send their data to memory-buffer boards or data-processing boards. After sending their data, they return to the standby state. The VME CPU can read data from the buffers and write the data into the storage after data for several thousands of events are accumulated in VME memories. This approach reduces the interrupt latency of the main VME CPU. In addition, as mentioned above, the TPC and PSA systems operate independently and their data are sent in parallel to each VME memory module. The quantity of data for one event in SMILE-II is similar to that for SMILE-I~\cite{matsuoka_2015}. In PSA, the ADC data for all pixels are recorded event-by-event to recover the precise total energy by considering the small leakage around the main hit peak below the threshold. In addition, the SMILE-II trigger rate in the balloon experiment is expected to increase to approximately 100-200 Hz at middle latitudes from $\sim$25 Hz for SMILE-I. In this case, the new DAQ system has to manage about several megabytes per second continuously (10 times the rate of SMILE-I).

For any Compton camera, the resolution of the energy of an incident gamma ray is determined by the combined resolutions of the forward and backward detectors. In addition, the resolution of the angular resolution measure (ARM), which is the accuracy with which the Compton-scattering angle is measured, is determined mainly by the energy resolution. Thus, the energy resolution of both the TPC and PSAs is quite important. For example, even a slight variation in gain of the multianode PMT corrupts the absorption position of a scattered gamma ray in the PSAs, 
and any variation in the TPC gain shifts the energy-loss rate, which seriously deteriorates the detection efficiency. Thus, calibration of the energy measurements of both the TPC and PSA modules is very important to ensure the proper performance of the ETCC. Thus, the TPC and PSA have to be calibrated every day. To do this, all calibrations were managed by the TCU. To calibrate the PSAs, the TCU generates an event trigger as soon as they receive a hit signal from the PSAs without waiting for the hit signal from the TPC. To calibrate the TPC, the TCU treats the hit signal from the FE2009bal chips as an event trigger and sends a ring-buffer-stop signal to the TPC readout board. The PSA readout module and the TPC readout board in both calibration modes work the same as in the ETCC mode. 

\section{Performance of the new data acquisition system}
\begin{figure}[t]
	\centering
	\includegraphics[width=1.0\linewidth]{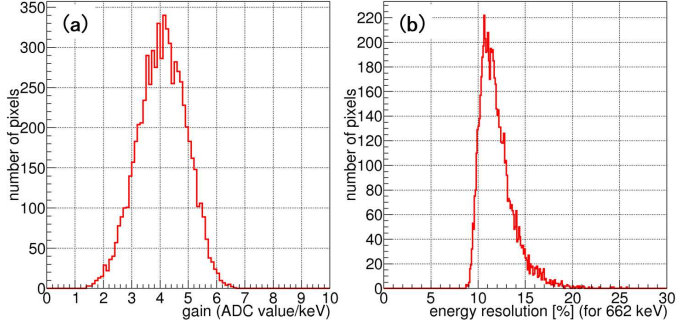}
	\caption{Distributions of (a) gain and (b) energy resolution for 6912 GSO pixel scintillators.
		\label{fig:psa_gain}
	}
\end{figure}
\begin{figure}[t]
	\centering
	\includegraphics[width=0.7\linewidth]{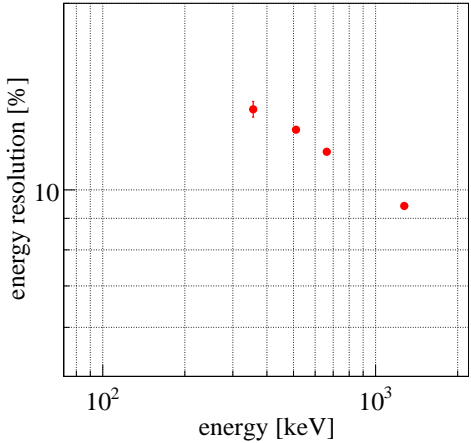}
	\caption{Energy resolution of entire scintillation camera as a function of incident energy.
		\label{fig:psa_e_res}
	}
\end{figure}
\begin{figure}[t]
	\centering
	\includegraphics[width=0.8\linewidth]{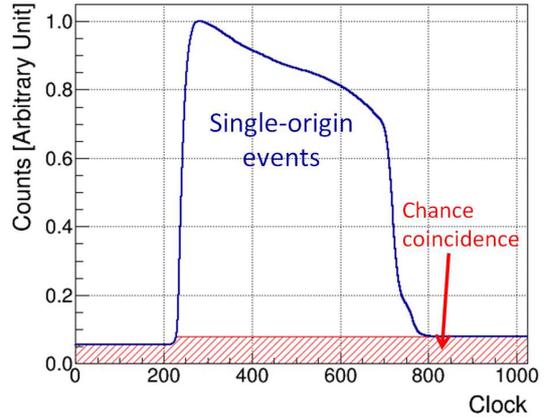}
	\caption{Clock-count distribution, which represents the time between the TPC-hit time and the ring-buffer-stop trigger. The clock count at the ring-buffer-stop trigger is 1023.
	  \label{fig:drift_clock}
	}
\end{figure}
\begin{figure}[t]
	\centering
	\includegraphics[width=0.8\linewidth]{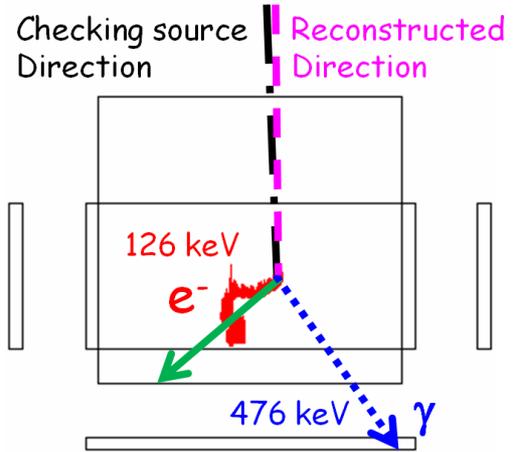}
	\caption{Typical gamma-ray event obtained by SMILE-II ETCC.
	  \label{fig:etcc_event}
	}
\end{figure}
\begin{figure}[t]
	\centering
	\includegraphics[width=0.8\linewidth]{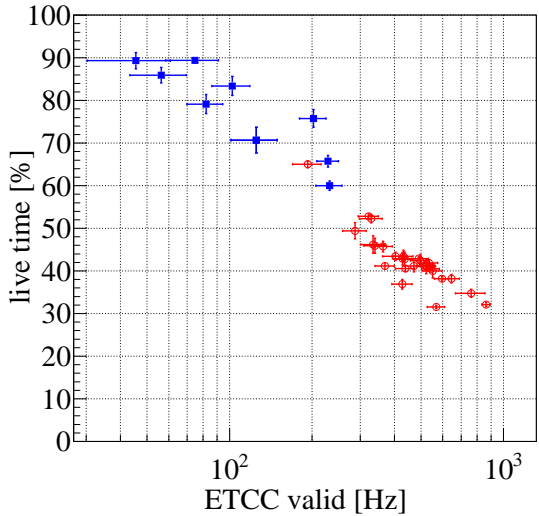}
	\caption{Variation of the live time of the DAQ as a function of the valid signal rate. The filled squares and open circles show data measured in the laboratory and RCNP, respectively.
	  \label{fig:daq_live_time}
	}
\end{figure}
The accuracy of the energy measurement of the scattered gamma-ray is important for all Compton cameras. However, the gain of the H8500 is not sufficiently uniform (the ratio of the maximum to the minimum for a single PMT is 2-3); hence, we must calibrate the energy response of PSAs for each pixel. The distribution of gain for all GSO pixel scintillators in the SMILE-II flight model ETCC is $\pm$25\% at 1$\sigma$, as shown in Fig.~\ref{fig:psa_gain}a. The energy resolution for each pixel is then distributed from 10\% to 14\% at 662 keV (Fig.~\ref{fig:psa_gain}b). Fig.~\ref{fig:psa_e_res} shows the energy resolution of the entire scintillation camera as a function of the incident energy. The energy resolution of the entire scintillation camera is approximately 11\% at 662 keV.

To confirm the performance of the new DAQ system, we operated the (30 cm)$^{3}$ ETCC with the ETCC mode by using the TCU, TPC readout boards, and the PSA readout modules. Fig.~\ref{fig:drift_clock} shows the distribution of the clock count, which represents the time between TPC-hit time and the TPC trigger. The timing at the stopping ring buffer is at 1023. Because the waiting time between the GSO trigger and the stopping ring buffer is 8 $\mu$s and the encoding clock is 100 MHz, the rising-edge timing of the GSO trigger is approximately 220. Therefore, the events distributed within the clock count of 230-750 are coincidence events between TPC and PSAs. The width of the coincidence distribution gives the maximum drift time; hence, we can measure the drift velocity at any time. In addition, we can estimate the signal-to-noise ratio by using this clock distribution, because the chance coincidence events are uniformly distributed. Upon irradiation by gamma rays from $^{137}$Cs, we searched for Compton-scattering events. The correlation between the energy deposited in the TPC and PSAs and the correlation between the track length and the gamma-ray energy deposited in the TPC are shown in Fig.~\ref{fig:etcc_spectra}a and \ref{fig:dedx_comp}b, respectively. Fig.~\ref{fig:etcc_event} shows a typical gamma-ray event obtained by ETCC upon irradiation by gamma rays from $^{137}$Cs.
More results are described by Mizumura et al.~\cite{mizumura_2014} and Tanimori et al.~\cite{tanimori_2015}.

To verify the background rejection and high counting rate of the SMILE-II ETCC, it was examined while exposed to intense radiation (140 MeV proton beam on a water target at Research Center for Nuclear Physics (RCNP), Osaka University). Brief reports have recently been published~\cite{matsuoka_2015,tanimori_2015}, and details of this experiment and the results for background noise rejection will be presented elsewhere. Here, a 140 MeV, 0.1-0.5 nA proton beam irradiates the 20 cm$\phi \times$30 cm cylindrical water target, from which similar amounts of fast neutrons and MeV gamma rays emanate. The ETCC was set normal to the beam line and at 1 m from the water target. A lead shield was placed between the water target and the ETCC to block the direct particles from the target, because in space, background particles uniformly irradiate the instrument. The trigger rate was varied from 100 Hz to 1 kHz. Fig.~\ref{fig:etcc_dEdx} shows the correlation between energy deposited and track length in the TPC at 450 Hz, in which the tangent of the curve gives the energy-loss rate dE/dx. Even under such condition of intense radiation, the cluster of Compton electrons halted in the TPC is clearly identified. This result certifies that SMILE-II operates stably even when exposed to about five times more radiation than expected in an actual balloon experiment. Fig.~\ref{fig:daq_live_time} shows the variation in the DAQ live time as a function of trigger rate, where the live time of 80\% at 100 Hz reduces to 20\% at 1 kHz. For actual measurements with the expected trigger rate of 100 Hz, a live time of 70\% is the minimum requirement which is satisfied with the present DAQ system. However, the margin for safety is small considering the considerable variation of trigger rates in the SMILE-II balloon experiment at 200 Hz, where the live time reduces to 60\%. At present, we use a relatively slow VME CPU, and data transfer via Ethernet is expected to increase the data-transfer rate more than several times without requiring a change of hardware.

\section{Summary}
We developed an ETCC technology for use in MeV gamma-ray astronomy and for applications such as nuclear-medicine gamma-ray imaging. In 2006, we did the first balloon-borne experiment SMILE-I and the small-size ETCC detected diffuse cosmic gamma rays and atmospheric gamma rays. As a next step, we developed the SMILE-II medium-sized ETCC with a (30 cm)$^{3}$ TPC and 108 PSAs. We plan to launch this instrument on a balloon within a few years to test its imaging and background noise rejection by observing bright sources, such as the Crab nebula. Because of the balloon-related requirements and the larger detection area, we also developed new readout electronics for both the TPC and PSAs as well as a new DAQ system. In addition, to increase the effective area by a factor of $\sim$100 compared with that of SMILE-I ETCC, we improved the data-acquisition algorithm to find the hit positions in the TPC efficiently and new readout electronics. These allow us to obtain clearer electron-track data as compared with SMILE-I. We also obtained satisfactory effective area, angular resolution, and the data-handling capability for the balloon experiment~\cite{tanimori_2015}. After SMILE-II, we plan to increase the effective area to 30 cm$^2$ at 300 keV (over 30 times larger than the effective area of the SMILE-II ETCC) by increasing the volume of the TPC to 40$\times$40$\times$40 cm$^3$ and the gas pressure up to 3 atm~\cite{tanimori_2015}. In this case, the expected trigger rate should also increase by a factor of 30; namely, to a few kHz. To manage such a high trigger rate, we are now developing a network data-acquisition system wherein all TPC and PSA readout boards are directly connected via Ethernet.

\section*{Acknowledgments}
This study was supported by a Grant-in-Aid for Scientific Research from the Ministry of Education, Culture, Sports, Science and Technology (MEXT) of Japan (Grant numbers 21224005, 20244026, 23654067, 25610042), a Grant-in-Aid from the Global COE program ``Next Generation Physics, Spun from Universality and Emergence'' from the MEXT of Japan, and a Grant-in-Aid for JSPS Fellows (Grant numbers 09J01029, 11J00606, 13J01213). This study was also supported by ``SENTAN'' program promoted by Japan Science and Technology Agency (JST). During the development of FE2009bal chips, this study was supported by VLSI Design and Education Center (VDEC), the University of Tokyo in collaboration with Cadence Design Systems, Inc. During the development of the $\mu$-PIC readout boards and the PMT high-voltage-supply boards, technical support was provided by KEK-DTP and Open-It (Open Source Consortium of Instrumentation)~\cite{open_it}.

\end{document}